# Collaborative Astronomical Image Mosaics

Daniel S. Katz, Computation Institute, University of Chicago & Argonne National Laboratory, d.katz@ieee.org

G. Bruce Berriman, Infrared Processing and Analysis Center, California Institute of Technology, gbb@ipac.caltech.edu

Robert G. Mann, Institute for Astronomy, University of Edinburgh, rgm@roe.ac.uk

## ABSTRACT

This chapter describes how astronomical imaging survey data have become a vital part of modern astronomy, how these data are archived and then served to the astronomical community through on-line data access portals. The Virtual Observatory, now under development, aims to make all these data accessible through a uniform set of interfaces. This chapter also describes the scientific need for one common image processing task, that of composing individual images into large scale mosaics and introduces Montage as a tool for this task. Montage, as distributed, can be used in four ways: as a single thread/process on a single CPU, in parallel using MPI to distribute similar tasks across a parallel computer, in parallel using grid tools (Pegasus/DAGMan) to distributed tasks across a grid, or in parallel using a script-driven approach (Swift). An on-request web based Montage service is available for users who do not need to build a local version. We also introduce some work on a new scripted version of Montage, which offers ease of customization for users. Then, we discuss various ideas where Web 2.0 technologies can help the Montage community.

## INTRODUCTION

Astronomy is moving from a discipline in which large numbers of practitioners collect and analyze their own data, to one where surveys collect data and make them accessible to the astronomical community for analysis. . This change is a consequence of improvements in detector sensitivity, the growth in the size of detector arrays, advances in automated telescope control, and increases in computing power to process the data. These data are served by archives and data centers, who curate and serve them through on-line data access portals. There has been much interest in developing uniform data access and data discovery services that can be used across all data sets. This is the principle of the Virtual Observatory. To support the goal of uniform data access, many nationally-funded Virtual Observatory projects formed the International Virtual Observatory Alliance (IVOA)[i] to develop the necessary standards on an international basis.

## ASTRONOMY AND DATA

Astronomers study the structure of the Universe by making measurements of the sky with ground-based or space-based telescopes. Observations from space are necessary at wavelengths absorbed by the Earth's atmosphere, and provide image stability and uninterrupted time coverage for projects such as detection of transiting planets by the Kepler[ii] mission.

Modern ground-based and space-based astronomical images are measured with cameras equipped with arrays of Charge Coupled Devices (CCDs), composed of individual pixel elements (pixels), usually cooled to cryogenic temperatures to maximize sensitivity. These devices convert radiation into electrical signals that can be converted into units of energy (flux) required for scientific analysis.

Rapid advances in the size and sensitivities of CCD cameras, along with advances in computational power and automated operation of instruments, have made this the age of the astronomical image survey. These surveys may cover a wide or small area, but generally operate in the same fashion. The survey images obtained at a telescope or in space are calibrated by a science team, and after an appropriate proprietary period, released to the public as science products. Calibrated in this sense implies that all instrumental signatures are removed from the data and any residual image artifacts are reported as such (e.g., defective pixels). The signals are converted to energy units, and the positions of the pixels on the sky are computed by reference to stars of known positions. Such images are ready for scientific analysis. Collections of images are often accompanied by catalogs of sources (stored as tables) extracted from the survey images by automated pipelines, optimized to take account of the effects of atmospheric distortion on the images and the resolving power of the instrument. These catalogs are served as science products, and are often more valuable in research than the images themselves. Additional data in some surveys are spectra associated with the sources in that survey, also stored in tables.

Perhaps the two most celebrated wide area image surveys are the Sloan Digital Sky Survey (SDSS, http://www.sdss.org) and the Two Micron All Sky Survey (2MASS, http://www.ipac.caltech.edu/2mass/). SDSS covered one quarter of the northern sky in five pass bands designated u (0.35 μm), g (0.47 μm), r (0.62 μm), i (0.75 μm) and z (0.89 μm). 2MASS covered the entire sky at three near-infrared wavelengths: J (1.25 μm), H (1.6 μm) and $K_s$ (2.2 μm). Yet there are many surveys that are proving important in astronomical research. We mention here just one important collection of surveys, those of the plane of our Galaxy. They have been undertaken at many wavelengths, from the visible through the infrared to radio wavelengths, and are transforming the studies of the global history of star formation and studies of the dynamics of the interstellar medium.

The data products from these surveys are generally archived at archives and data centers chartered specifically for that purpose. Their aim is not simply to preserve the data, but to preserve and document technical knowledge about them that are essential to accurate scientific exploitation. Usually one archive will be charged with guaranteeing long-term availability of the data set, but copies are commonly held at several archives.

The data are discovered through queries on the attributes (metadata) describing the images. An example might be: "return all 2MASS images at J, H and K within a 5 x 5 degree box centered on the M31 Galaxy." The metadata are usually stored in relational databases designed to support these searches. Queries are made through a web form in a browser, or through a program interface that is embedded in a script. The queries return a table of images that meet the search criteria. The tables will contain all the metadata for each image, and a handle that describes the location of each image. Services to perform further filtering of the results are often provided, as are services to package and deliver the images to the astronomer.

The worldwide Virtual Observatory projects are developing a series of program interfaces that, when complete, will support common queries to diverse data sets across many distributed archives. The query results will be transferred in a self-describing, machine-independent XML structure called VOTable. Tools to reformat VOTable into astronomer-friendly formats such as ASCII tables will be available.

The images are stored in files that adhere to the definition of the Flexible Image Transport System (FITS) standard[iii]. FITS is the format adopted by the astronomical community for data interchange and archival storage. Briefly, FITS is a data format designed to provide a platform-independent means for exchange of astronomical data. A FITS data file is composed of a fixed logical record length of 2880 bytes. The file can contain an unlimited number of header records (or metadata), 80 bytes long, having a '*keyword=value*' format. These headers describe the organization of the binary image data, the format of the contents, and the attributes of the data (telescope, observation date, filter, coordinates, etc.) that are common over all the binary data in the file. The keywords can be customized according to needs and provides the flexibility to provide a full description of the image, including quality information, information on residual defects in the image and a processing history. An example of the headers in a FITS file is shown in Figure 1.

The headers are followed by the image data, always represented as a two-dimensional binary array. The data array is a two-dimensional representation of the three-dimensional curved surface of the sky, where the coordinates of the array are right ascension and declination, and the value of the elements of the array is energy measured at those coordinates. By analogy with terrestrial cartography, a mathematical projection describes the relationship between the pixel coordinates in the image and the pixels on three-dimensional surface of the sky. As in cartography, there are many projections in use in astronomy. The World Coordinate System (WCS)[iii] formally describes the relationships between pixel representations and physical units on the sky. The WCS includes a definition of how celestial coordinates and projections are represented in the FITS format as *keyword=value* pairs in the file headers.

```
ORDATE  = '000503  '          / Observation Ref Date (yymmdd)
DAYNUM  = '1160    '          / Observation Day Num
FN_PRFX = 'j1160059'          / .rdo and .par filename prefix
TYPE    = 'sci     '          / Scan type: dar flt sci cal tst
SCANNO  =                 59  / Scan Number
SCANDIR = 'n       '          / Scan Direction: n, s, -
COMMENT                          (OV)
STRIP_ID=             301788  / Strip ID (OV)
POSITNID= 's001422 '          / Position ID (OV)
ORIGIN  = '2MASS   '          / 2MASS Survey Camera
CTYPE1  = 'RA---SIN'          / Orthographic Projection
CTYPE2  = 'DEC--SIN'          / Orthographic Projection
CRPIX1  =              256.5 /   Axis 1 Reference Pixel
CRPIX2  =              512.5 /   Axis 2 Reference Pixel
CRVAL1  =        215.6251831 /   RA  at Frame Center, J2000 (deg)
CRVAL2  =      -0.4748106667 /   Dec at Frame Center, J2000 (deg)
CROTA2  =     1.900065243E-05 /   Image Twist +AXIS2 W of N, J2000 (deg)
CDELT1  =   -0.0002777777845 /   Axis 1 Pixel Size (degs)
CDELT2  =    0.0002777777845 /   Axis 2 Pixel Size (degs)
USXREF  =             -256.5 /   U-scan X at Grid (0,0)
USYREF  =             19556. /   U-scan Y at Grid (0,0)
```

**Figure 1.** A sample of the metadata describing a 2MASS image J-band image, written in the form of keyword=value pairs, in compliance with the definition of the Flexible Image Transport System (FITS).

## MONTAGE

An image mosaic is a combination of individual pixel data in many images so that the data appear to be from a single image measured with a telescope or spacecraft. There are a variety of reasons why astronomers are interested in building mosaics, including studying structures in the sky that are larger than individual images, to perform multi-wavelength image federation, and to build high signal-to-noise images for studies of faint sources. In the first case, the astronomer wants to see the full resolution of the image, but wants to see multiple images at that resolution as if they were one image. In the second case, images are invariably grayscale, where the brightness of the pixels corresponds to the flux measured by the CCD at those coordinates. An astronomer can combine multiple images, which contain multiple wavelengths, by using color – for example, using red for one wavelength's image, blue for another, and green for a third. If the images were taken from different instruments, they must be processed so that they all have the same pixel pattern on the sky (re-projection), have the same spatial sampling on the sky, and are represented in a common coordinate system. The sky background radiation must be rectified to a common level across all the images, and then all the processed images must be added to make the final mosaic.

From 2002 to 2005, a team at IPAC, the Center for Advanced Computing Research (CACR) and the Jet Propulsion Laboratory designed and built a toolkit called Montage[iv], designed to deliver science grade astronomical image mosaics for these reasons. Montage was designed to preserve astrometry (position) and photometry (flux) of images, and rectify backgrounds to a common level (needed because of varying sky emission, instrument signatures, etc.) and then co-add all the images.

Montage delivers custom mosaics, where the user specifies the input data, and for the output mosaic, the projection, coordinates, spatial sampling, mosaic size, and image rotation. It uses an extensible "toolkit" design, with stand-alone (loosely coupled) engines for image reprojection, background rectification, and co-addition. This provides flexibility to users; e.g., one can use Montage only as a reprojection and co-registration engine, without doing any co-addition. Or one can apply custom algorithms for parts of the processing, e.g., co-addition or background rectification, with no impact on other engines or the overall flow of Montage.

Montage has been implemented in ANSI C for portability. Each tool is an executable, with inputs specified on the command line and through files, and outputs as files. The toolbox design lowers testing costs, as the team can perform independent tests of all modules as well as a small number of system tests, rather than a very large set of systematic testing, as would be needed if Montage was a monolithic application.

Three different versions of Montage have been built. The first version focused on the basic requirements: functionality and scientific accuracy. It was able to run on a single compute core, with overall speed a much lower goal than accuracy. The second version improved performance in a number of regards, including reducing the amount of memory required for the final coaddition step, improving the speed of the reprojection algorithm by using a faster algorithm[v] and by taking advantage of the inherent parallelism in Montage on both parallel computing and grid computing infrastructures[vi]. The third version of Montage increased usability, including adding data access modules, the ability to create tiled output, a tool to build multi-band jpeg images from FITS images, and other improvements in speed and accuracy. Of course, each version also included bug fixes discovered in the previous version.

Montage can be downloaded and used locally, or used on a grid or parallel system. In addition, IPAC supports Montage as a public service, where users order mosaics through a web portal. This uses a small cluster at IPAC for small jobs, and the service uses public resources, such as TeraGrid[vii] and EC2[viii], for larger jobs.

Montage is widely used in the astronomical community to support astronomical research, generate data products for dissemination to the community, perform quality assurance of data products, and develop of on-line image access and discovery services[ix]. Three examples of this are: 1) Stock and Barlow[x] used Montage to support their search for ejecta from WR-stars by creating mosaics of images from the AAO/UKST Southern Hα Survey (SHS). 2) Anderson et al.[xi] used Montage to resample 350 μm data from the Herschel Space Telescope in their study of the physical properties of the dust of the HII region RCW 120. 3) As shown in Figure 2, the Arecibo Legacy Fast ALFA survey[xii] has used Montage to create a large wide-field science-grade mosaic of HI emission at high galactic latitude. The survey will cover 7000 square degrees of sky with the 305 meter Arecibo Radio Telescope (http://www.naic.edu). When complete, it will be the largest collection of 21-cm neutral hydrogen (HI) catalogs, images, and high quality radio spectra ever obtained by astronomers.

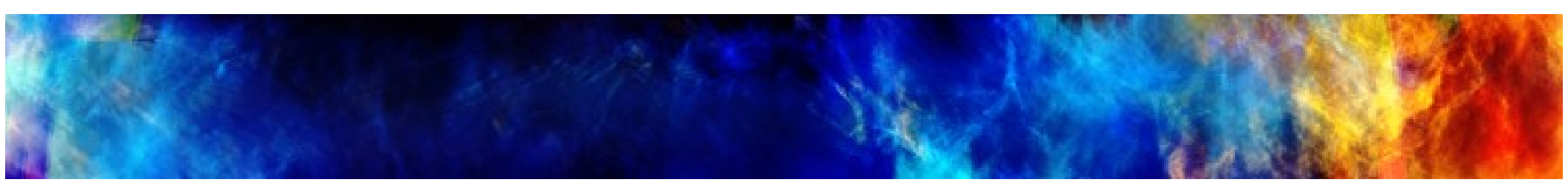

**Figure 2.** 1500 square degree equal area Aitoff projection mosaic, computed with Montage, of galactic neutral hydrogen observed with the ALFALFA survey near the North Galactic Pole (NGP). Image courtesy of Dr. Brian Kent and the ALFALFA collaboration.

The generality of Montage as a workflow application has led it to become an exemplar for computer scientists who study workflows and workflow-based applications, such as those working on Pegasus[xiii], ASKALON[xiv], QoS-enabled GridFTP[xv], SWIFT[xvi], SCALEA-G[xvii], VGRaDS[xviii], etc., and investigations are also taking place into the use of Montage on clouds[xix].

Some key Montage components are:

- mImgtbl; extracts geometry information from a set of FITS headers and create a metadata table from it.
- mProject; reprojects a FITS image.
- mProjExec; runs mProject for each image in an image metadata table.
- mOverlaps; analyzes an image metadata table to determine which images overlap on the sky.

- mDiff; performs a simple image difference between a pair of overlapping images (in the same projection).
- mDiffExec; runs mDiff on all the overlap pairs identified by mOverlaps.
- mFitplane; fits a plane (excluding outlier pixels) to an image. Meant for use on the difference images generated by mDiff.
- mFitExec; runs mFitplane on all mOverlaps pairs. Creates a table of image-to-image difference parameters.
- mBgModel; uses an image-to-image difference parameter table to interactively determine a set of corrections to apply to each image to achieve a "best" global fit.
- mBackground; applies a planar correction to an image.
- mBgExec; runs mBackground on all images in an image metadata table
- mAdd; coadds the set of images to produce an output mosaic.

Figure 3 shows an example of how these modules can be used to create a mosaic.

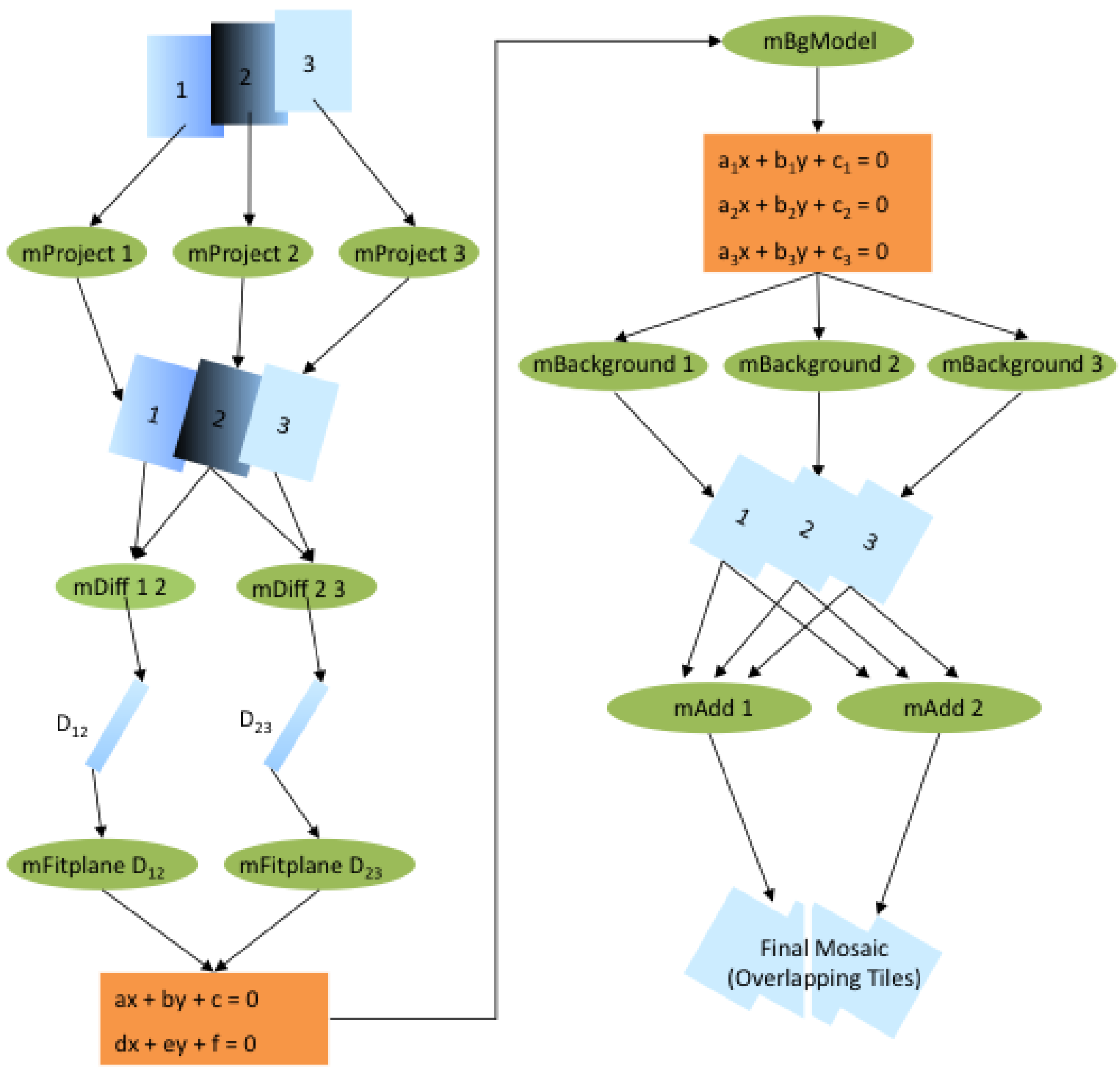

**Figure 3.** Montage workflow, showing three input images being reprojected, having background rectification calculated and applied, and being co-added into a two-tiled mosaic.

The Montage modules can be run directly as shown in Figure 3, which works on any computer system. In addition, the use of the modules can be adapted in three ways, to take advantage of parallel or grid computers.

First, Montage can be built with MPI[xx]. Each *Exec routine can be replaced by a *ExecMPI routine, which is built by changing a flag in the Montage build script (makefile). The MPI executives are parallelized straightforwardly, with all processes of a given executive being identical to each other. All the initialization is duplicated by all of the processes. A line is added at the start of the main loop, so that each process only calls the sub-module if the remainder of the loop count divided by the number of processes equals the MPI rank (a logical identifier of an MPI process). All processes then participate in global sums to find the total statistics of how many sub-modules succeeded, failed, etc., as each process keeps track of its own statistics. After the global sums, only the process with rank 0 prints the global statistics. In addition, Montage includes an MPI version of mAdd, which is somewhat more complicated. In it, each process has a unique section of the output mosaic it "owns", and each process reads all images that might contribute to that section, does the co-addition for its section, and then writes out that section into a common mosaic file. To build a mosaic using the MPI version of Montage, the user can type (or use a shell script) the same commands that were used in the sequential code, simply replacing each *Exec execution with a call to mpirun to run the *ExecMPI version.

The MPI version of Montage has very good parallel performance for routines that do a good amount of computing, but suffers from I/O contention where multiple processes are reading or writing the same files. For a set of processors that share a common file system with good I/O performance, the MPI version is probably the best way to run a single Montage instance quickly. However, if any one of those processors fails, the entire Montage job will fail and will have to be restarted from the *ExecMPI routine that failed.

A second approach at taking advantage of the parallelism in Montage comes from recognizing that the Montage tasks (the ovals in Figure 3) can also be run in a different order than simply stage by stage. For example, after mProject 1 and mProject 2 are run, mDiff 1 2 can be run next. Here, we capture the Montage dependencies as a directed acyclic graph (DAG). Montage is set up to build this DAG for a given set of data and a specified flow of processing using a module called mDAG. This DAG is an abstract DAG, as it contains the flow of files and processing, but does not specify where this processing should be done. A tool called Pegasus[xxi] can be used to map the abstract DAG to a concrete DAG that includes where the processing should be done, and any needed data transfer steps, and then another tool called DAGMan[xxii] can execute this DAG. We refer to this series of work/tools as the grid version of Montage.

The grid version generally runs slightly slower (5-10%) than the MPI version on the same set of processors, partially because it has a different set of work to do, including running the serial mDAG module to build the DAG, and the serial Pegasus tool to map it to processors. However, it also has some benefits over the MPI version, including the fact that it can run on multiple sets of processors that do not share a common disk, and that the fault tolerance is finer-grained; the DAG can be restarted from a failure, but more often simply the failed task can be rerun on a different processor.

Finally, recent work on Montage has included investigating a scripting approach, using Swift[xxiii]. Scripting allows the user to easily change the flow of processing in Montage, including removing tools from the standard flow, and adding customized tools. Using the Pegasus/DAGMan grid version of Montage, this cannot easily be done, as it would be relatively hard for a user to change the Montage-supplied mDAG code that builds the initial DAG used to create the workflow that is the main input to Pegasus. The workflow of the MPI version is easier to change, as it is just a script, but a user who writes a new module would also have to write an MPI version of that module, which would be fairly hard unless the user happens to be an experienced MPI programmer. The input to the Swift version of Montage is a fairly simple and concise script that uses a foreach statement to perform a task over a set of files, given a single processor tool to execute that task. Swift also optimizes the execution of the tasks to be executed over a set of processors that could be on a supercomputer or a grid, similar to Pegasus.

### WEB 2.0

As the ideas behind Web 2.0 have been discussed earlier in this book, here we discuss some ideas of how Web 2.0 can be applied to the building a user community and scientific projects such as building collaborative astronomical image mosaics.

One simple idea that is already being used is a wiki/blog for discussion of how to use Montage, successes and failures that users have, and user feedback that can be used to improve Montage. The interactive website Astrobetter (http://www.astrobetter.com/) offers astronomers the opportunity to share tools and expertise, and image mosaic creation is one of the topics that has been discussed on this site[xxiv]. The comments have in fact led directly to improvements in the organization and content of the Montage web site.  They are also leading Montage to deploy a dedicated user forum for users to share innovative uses of the Montage toolkit, workarounds to problems and identify defects.

Scientists in many fields have generally not taken to Web 2.0 as a collaborative tool and as a means of sharing data. A research report by the Research Information Network, a British policy unit, is to our knowledge the only detailed study of why this is the case in the fields of social, physical and biological sciences[xxv]. The primary barrier is apparently a lack of understanding of how to get started and what benefits will accrue to scientists (See also the blogpost by Bruce Berriman on this issue, "25 Things for Researchers and Social Media!"[xxvi]. Despite this lack of take-up, there are many science and outreach projects that can be done effectively with Web 2.0. Some examples are described below; they illustrate the unrealized potential of Web 2.0 in astronomy, a discipline increasingly dominated by large, multi-institutional (and, often, multi-national) research consortia, which in theory should make it an ideal domain for Web 2.0.

One project would be building a collection of image mosaics produced by many users. This could be done at many levels. For example, the Montage project could host mosaics produced by others. This seems particularly useful in the case of images that are produced for outreach. Simple outreach mosaics (or science mosaics) could also be placed on Flickr. Or users could post their own mosaics on servers that they choose. In any of these cases, searching or browsing of these images also need to be considered. If all images were on one site, they would be easy to browse. If they are on multiple sites, a registry (or a federation of registries) is needed. Additionally, image tags are needed for searching. FITS files already have metadata incorporated, but the standardization of this metadata for images is could be handled by a standard that is under development for this purpose[xxvii]. This tagging could be manual or automated. The registry could also have an RSS feed or use twitter to allow interested groups to see new mosaics that had been registered.

Web 2.0 also potentially could allow collaborative building of atlases from surveys. For example, using the virtual data concept[xxviii], once an atlas is defined (meaning a definition of a set of image plates from one or more survey bands that would be useful to a community of users, including a projection of the plates, and the boundaries of the individual plates), a library could be built with pages for each plate. However, each plate would not be built until a user wanted it. A request for a plate that had not yet been built would launch a run of Montage. This could either be done for public data, with the atlas being publicly accessible, or it could be done only within a collaboration, with only the collaborators having access to the atlas initially. This could be done by Montage or others, or it could be done as a service that would provide overlays over Google Sky, for example. Having such atlases available might be helpful to researchers who want to compare new data against some data of known quality and provenance.

Using the mashup concept, Montage components could be turned into services that could be exposed to users with inputs and outputs named to make them clearly understandable, perhaps as Google gadgets. This would require the allow users to compose services as they choose, including mashing them up with other non-Montage services, though the cost of running these services would have to borne by someone, perhaps Montage or an infrastructure such as a TeraGrid science gateway or a hosted commercial service such as Amazon EC2 or Windows Azure.

By the early 2000s, many scientists were designing unique user interfaces and tools to access compute and data resources. The TeraGrid Science Gateways program[xxix] started in late 2004 to try to encourage this to continue, with common tools being developed for the scientists and communities who were building the gateways, and TeraGrid resources being used for the compute and storage parts. This was originally done using mostly Web 1.0 technologies, but currently, the TeraGrid Science Gateways program is investigating how to integrate Web 2.0 technologies as well.

### A Web 2.0 Scenario

Some of these ideas can be explored in a scenario that illustrates how Montage-based services could enhance collaboration within a fictitious research consortium. Similar to many in contemporary astronomy, this consortium comprises researchers from more than a dozen institutions from a handful of different countries. Its focus is a large-area survey being undertaken by an equally fictitious satellite mission that observes in the far-infrared region of the electromagnetic spectrum. In order to understand the astrophysical nature of the sources detected in their far-infrared survey, the consortium is collating existing data in their survey field to be found in data archives published through the Virtual Observatory, as well as undertaking its own follow-up observations.

Annie coordinates the team (within the consortium) collecting data from the visible region of the spectrum. She has collected together information relating to all existing image archives accessible via the Virtual Observatory (VO) that have data covering the consortium's survey field, and has led an observing programme plugging the gaps in the coverage of the survey field in all five of the SDSS bands. She has previously run all these data through Montage to produce a homogeneous atlas of the survey field. In fully materialized form, this atlas takes up a lot of disk space, so it current exists only virtually, as a list of file URIs and associated configuration and calibration parameters.

Annie's colleague, Barney, is working on the far-infrared survey data, and releases a new map and associated source list to the consortium. Annie quickly sends the source list to a Montage-based service, which generates a low-resolution version of her sky atlas, suitable for exploration through a browser, using an image display tool similar to Google Sky. The service also generates a KML overlay showing the boundary of the far-infrared map, and the position (with error region) of each source in Barney's list. Annie announces the availability of this by circulating a URI to the consortium, and it is picked up by Charlie and Denise, two other members of the consortium. Charlie has been mapping the survey field in the radio, so he feeds the Montage KML-generation service with the positions of his radio sources, and they are published as an addendum to Annie's original overlay, so that consortium members can see which radio and far-infrared sources are coincident and what they look like in the optical. Denise looks at this enhanced overlay, and sees that one of the sources in Barney's list is very bright in the far-infrared, but appears to have no counterpart in Charlie's radio source list. She zooms into that position (with the Montage-based service generating a higher-resolution image on-the-fly as she does) and sees a very faint blob, which is not included in the source catalogues Annie generated from her images. Denise interprets this as indicating that the source is too faint for inclusion in Annie's catalogues, but it is clearly real, as it appears in several different visible bands, so she defines an aperture based on the position of Barney's far-infrared source and the image display tool measures the flux within it in each of the optical bands present in Annie's image atlas for the field, and an upper limit to the radio flux, through calls to a flux measurement routine that is executed on the Montage server (to which Charlie's radio map has also been uploaded).

Denise then adds an annotation to the overlay tag for this source in Barney's list, noting these new flux data and explaining briefly what she did. Meanwhile, Ernie, Felicity and George are, similarly, browsing the growing image-plus-overlays dataset, and make further annotations, recording possible associations with other existing datasets, noting apparent problems with Barney's far-infrared map and, together, helping build up the consortium's multi-wavelength view of their survey field.

## CONCLUSION

Astronomical imaging survey data have become a vital part of modern astronomy. One common image processing task that is done on such data is composing individual images into large scale mosaics, which can be done with Montage. Montage has been downloaded by over 4,000 users, including astronomers and computer scientists, for whom it has become one of the most-studied workflow applications.

There are at least four versions of Montage that have been developed (at least in part) by the authors, and each has certain advantages and disadvantages on a set of parallel and distributed platforms.

Astronomers, like many physical scientists, have not taken to Web 2.0 as a science collaborative tool. Yet the science and outreach applications of Web 2.0 identified here certainly reveal its potential. Most of the constituent parts of the Web 2.0 scenario outlined above can already be implemented using VO-enabled tools, but what is lacking is the ability to share annotations on a multi-resolution representation of an image dataset, capable of being explored through simple zoom and pan operations. Even for astronomers used to integrating multi-wavelength datasets, there remains something special about visible image data, and services like those outlined above could significantly enhance the way that research consortia develop a collaborative view of their data.

## ACKNOWLEDGEMENTS

G. B. B. is supported by the NASA Exoplanet Science Institute at the Infrared Processing and Analysis Center, operated by the California Institute of Technology in coordination with the Jet Propulsion Laboratory (JPL).

Montage was funded by the National Aeronautics and Space Administration's Earth Science Technology Office, Computation Technologies Project, under Cooperative Agreement Number NCC5-626 between NASA and the California Institute of Technology. Montage is maintained at the NASA/IPAC Infrared Science Archive.

## REFERENCES

[i] International Virtual Observatory Alliance, http://www.ivoa.net/
[ii] Kepler Mission, http://kepler.nasa.gov/
[iii] Calabretta, M. R. and E. W. Greisen. Representations of celestial coordinates in FITS. *Astronomy and Astrophysics* **395**:1077-1122, 2002.
[iv] Berriman, G. B., D. Curkendall, J. Good, J. Jacob, D. S. Katz, M. Kong, S. Monkewitz, R. Moore, T. Prince, and R. Williams. An Architecture for Access to a Compute Intensive Image Mosaic Service in the NVO. *SPIE Conference 4846: Astronomical Telescopes and Instrumentation*, 2002.
[v] Makovoz, D. and I. Khan. Mosaicking with MOPEX. *Proceedings of ADASS XIV*, 2004.
[vi] Jacob, J. C., D. S. Katz, G. B. Berriman, J. C. Good, A. C. Laity, E. Deelman, C. Kesselman, G. Sing, M.-H. Su, T. A. Prince, and R. Williams. Montage: A Grid Portal and Software Toolkit for Science-Grade Astronomical Image Mosaicking. *International Journal of Computational Science and Engineering* **4**(2):73-87, 2009.
[vii] Catlett, C. "The Philosophy of TeraGrid: Building an Open, Extensible, Distributed TeraScale Facility," *Proceedings of 2nd IEEE/ACM International Symposium on Cluster Computing and the Grid*, 2002.
[viii] Amazon Elastic Compute Cloud (Amazon EC2), http://aws.amazon.com/ec2/
[ix] Applications of Montage, http://montage.ipac.caltech.edu/applications_data.html
[x] Stock, D. J., M. J. Barlow. A search for Ejecta Nebulae around Wolf-Rayet Stars using the SHS Hα survey, accepted for publication in *Monthly Notices of the Royal Astronomical Society*, 2010. arXiv:1006.0625 [astro-ph.GA]
[xi] Anderson, L. D., A. Zavagno, J.A. Rodon, D. Russeil, A. Abergel, P. Ade, P. Andre, H. Arab, J.-P. Baluteau, J.-P. Bernard, K. Blagrave, F. Boulanger, M. Cohen, M. Compiegne, P. Cox, E. Dartois, G. Davis, R. Emery, T. Fulton, C. Gry, E. Habart, M. Huang, C. Joblin, S.C. Jones, J. Kirk, G. Lagache, T. Lim, S. Madden, G. Makiwa, P. Martin, M.-A. Miville-Deschenes, S. Molinari, H.

Moseley, F. Motte, D.A. Naylor, K. Okumura, D. Pinheiro Gocalvez, E. Polehampton, P. Saraceno, S. Sidher, L. Spencer, B. Swinyard, D. Ward-Thompson, G.J. White 2010. The physical properties of the dust in the RCW 120 HII region as seen by Herschel. accepted for publication in *Astronomy & Astrophysics*, 2010. arXiv:1005.1565 [astro-ph.GA]

[xii] The Arecibo Legacy Fast ALFA Survey, http://egg.astro.cornell.edu/

[xiii] Deelman, E., J. Blythe, Y. Gil, C. Kesselman, G. Mehta, and K. Vahi. Mapping Abstract Complex Workflows onto Grid Environments. *Journal of Grid Computing*, **1**(1), 2003.

[xiv] Wieczorek, M., R. Prodan, and T. Fahringer. Scheduling of Scientific Workflows in the ASKALON Grid Environment. *ACM SIGMOD Record*, **34**(3):52-62, 2005.

[xv] Humphrey, M. and S.-M. Park. Data Throttling for Data-Intensive Workflows. *Proceedings of 22nd IEEE International Parallel and Distributed Processing Symposium*, 2008.

[xvi] Sotomayor, B., K. Keahey, I. Foster, and T. Freeman. Enabling Cost-effective Resource Leases with Virtual Machines. *Proceedings of HPDC*, 2007.

[xvii] Truong, H.-L., T. Fahringer, and S. Dustda. Dynamic Instrumentation, Performance Monitoring and Analysis of Grid Scientific Workflows. *Journal of Grid Computing.* **2005**(3):1-18, 2005.

[xviii] VGRaDS: Montage, a Project Providing a Portable, Compute-Intensive Service Delivering Custom Mosaics on Demand. http://vgrads.rice.edu/research/applications/montage.

[xix] Juve, G., E. Deelman, K. Vahi, G. Mehta, B. Berriman, B. Berman, and P. Maechling. Scientific Workflow Applications on Amazon EC2. *Proceedings of IEEE International Conference on e-Science*, 2009.

[xx] Snir, M., S. W. Otto, S. Huss-Lederman, D. W. Walker, and J. Dongarra, *MPI: The Complete Reference*. The MIT Press, 1996.

[xxi] Deelman, E., J. Blythe, Y. Gil, C. Kesselman, G. Mehta, K. Vahi, K. Blackburn, A. Lazzarini, A. Arbree, R. Cavanaugh, and S. Koranda. (2003) Mapping abstract complex workflows onto grid environments, *Journal of Grid Computing*, **1**(1):25-39. 2003.

[xxii] Frey, J., T. Tannenbaum, M. Livny, and S. Tuecke. Condor-G: a computation management agent for multiinstitutional grids, *Proceedings of the 10th IEEE International Symposium on High-Performance Distributed Computing*, 2001.

[xxiii] Zhao, Y., M. Hategan, B. Clifford, I. Foster, G. von Laszewski, V. Nefedova, I. Raicu, T. Stef-Praun, and M. Wilde.. Swift: Fast, Reliable, Loosely Coupled Parallel Computation. *Proceeedings of 1st IEEE International Workshop on Scientific Workflows*, 199-206, 2007.

[xxiv] Astrobetter web site, Better Ways to Make Large Image Mosaics discussion thread, http://www.astrobetter.com/better-ways-to-make-large-image-mosiacs/

[xxv] Research Information Network. If you build it, will they come? How researchers perceive and use web 2.0. A Research Information Network report, http://www.rin.ac.uk/web-20-researchers, July 2010.

[xxvi] Berriman, G. B. 25 Things for Researchers and Social Media!. https://astrocompute.wordpress.com/2010/09/15/25-things-for-researchers-and-social-media/, September 15, 2010.

[xxvii] Astronomy Visualization Metadata Standard, http://www.virtualastronomy.org/avm_metadata.php

[xxviii] Foster, I., J. Voeckler, M. Wilde, and Y. Zhao. Chimera: A Virtual Data System for Representing, Querying, and Automating Data Derivation. *Proceedings of 14th International Conference on Scientific and Statistical Database Management (SSDBM'02)*, 2002.

[xxix] Wilkins-Diehr, N., D. Gannon, G. Klimeck, S. Oster, and S. Pamidighantam. TeraGrid science gateways and their impact on science. *Computer*, **41**(11):32-41, 2008.